\documentclass[useAMS,usenatbib]{mn2e}
 \usepackage{graphicx}
 \usepackage[letterpaper]{geometry}
 \usepackage{bm}
 \usepackage{amsmath}
 \usepackage{amssymb}

\def\lap{\hbox{${_{\displaystyle<}\atop^{\displaystyle\sim}}$}}
\def\gap{\hbox{${_{\displaystyle>}\atop^{\displaystyle\sim}}$}}

\author[S. Price et al.]
  {S.~Price$^1$, B.~Link$^1$, S.N.~Shore$^2$, D.J.~Nice$^3$\\
             \\
  $^1$Department of Physics, Montana State University, Bozeman, MT 59717, USA\\
  $^2$Dipartimento di Fisica ÒEnrico FermiÓ and INFN - Sezione di Pisa, Universita` di Pisa, largo B. Pontecorvo 3, I-56127 Pisa, Italy\\
  $^3$ Physics Department, Lafayette College, Easton, PA 18042, USA}

\title{Time-Correlated Structure in Spin Fluctuations in Pulsars}

\date{\today}

\begin{document}

\onecolumn

\maketitle

\begin{abstract}
We study statistical properties of stochastic variations in pulse
arrival times, {\em timing noise}, in radio pulsars using a new
analysis method applied in the time domain. The method proceeds in two
steps. First, we subtract low-frequency wander using a high-pass
filter. Second, we calculate the discrete correlation function of the
filtered data. As a complementary method for measuring correlations,
we introduce a statistic that measures the dispersion of the data with
respect to the data translated in time. The analysis methods presented
here are robust and of general usefulness for studying arrival time
variations over timescales approaching the average sampling interval.
We apply these methods to timing data for 32 pulsars. In two radio
pulsars, PSRs B1133+16 and B1933+16, we find that fluctuations in
arrival times are correlated over timescales of $ 10-20$ d with the
distinct signature of a relaxation process. Though this relaxation
response could be magnetospheric in origin, we argue that damping
between the neutron star crust and interior liquid is a more likely
explanation. Under this interpretation, our results provide the first
evidence independent from pulsar spin glitches of differential
rotation in neutron stars.  PSR B0950+08, shows evidence for
quasi-periodic oscillations that could be related to mode switching.

\end{abstract}

\section{Introduction}

Radio pulsars are superb clocks. The regularity of the arrival times of
the pulsed emission, upon barycentering, subtraction of proper motion,
and gradual spin down, rivals atomic clocks in stability. Perhaps even
more interesting are the {\em imperfections} of pulsars as clocks;
variation in pulse arrival times, {\em timing noise}, is present at
some level in all pulsars, and has remained unexplained since the
discovery of pulsars in 1967.  Timing noise is a very complex process,
and there appear to be many contributing factors; \cite{hobbs_etal10} and
\cite{Cordes_Downs85} have shown that the complexity of
timing noise cannot be explained by high frequency random walks in the
pulsar spin parameters (see, for example, \citealt{boynton_etal72,cordes80}). 
Possible contributors to timing noise include
variable coupling between the crust and the liquid interior
\citep{alpar_etal86,jones90b}, stochastic adjustments to the star's
figure \citep{cordes93}, variations in the external electromagnetic
spin-down torque on the star \citep{cheng87a,cheng87b,ulw06,lyne10},
and time variability of the interstellar medium between the pulsar and
Earth \citep{liu_et_al11}. 

A neutron star is a dynamic system consisting of a rigid crust, a
liquid interior, and an active magnetosphere.  Variations in the spin
rate of the crust are the result of driving torques on the crust,
filtered by the response of the system to those torques. Generally in
noisy systems it is possible to determine properties of the response
function without knowledge of the forcing function. For example,
thermal fluctuations of the current in a circuit containing a resistor
of resistance $R$ in series with a capacitor of capacitance $C$ can be
used to measure the circuit's intrinsic decay time $RC$, independent of the
processes that drive the current fluctuations. In this paper we
develop techniques with which to measure properties of the response
function of the neutron star system. 

Using timing noise to probe the neutron star system is an old idea.
For a neutron star with a damped rotational mode
associated with, for example, friction between the crust and a portion
of the liquid interior, stochastic perturbations from rotational
equilibrium by the noise process would never relax completely. In this
case, decoupling by fluctuations at frequencies higher than the
damping frequency $\tau_d^{-1}$ would increase the spectral power for
all frequencies above $\tau_d^{-1}$ by a magnitude determined by the
ratio of the moments of inertia of the crust and the liquid to which it is
imperfectly coupled
\citep{lps78}. Past studies of timing noise power spectra have not
revealed structure beyond the power-law that is the hallmark of a
noisy process, that is, no deviations from rigid-body rotation have
yet been detected \citep{bd79,boynton81,boynton_etal84}. The only
evidence to date that neutron stars do not rotate rigidly comes from a
very different phenomenon than timing noise: the {\em glitches},
sudden increases in spin rate (see, for example, \citealt{lss00}),
whose occurrence and subsequent recovery have been attributed to
variable coupling between the crust and the liquid interior (see, {\sl
e.g.}, \citealt{alpar_etal84,leb93,pizz11}). Why neutron stars have
not shown signatures of deviations from rigid-body rotation in their
noise spectra has been an important open question in neutron star
physics for over three decades.

Since the early work on timing noise, the quantity and quality of data
have increased. In the largest and most comprehensive study of timing
noise to date, \cite{hobbs_etal10} presented timing residuals of 366
pulsars over nearly 40 years, some with nearly daily monitoring,
mainly by the Lovell Telescope at Jodrell Bank. Here we revisit the
issue of using timing noise to identify properties of neutron star
response with these high-resolution data using methods applied in the
time domain to 32 radio pulsars.  In two radio pulsars, PSRs B1133+16
and B1933+16, we find that fluctuations in arrival times are
correlated over timescales of $ 10-20$ d with the distinct signature
of a relaxation process.  Though the relaxation response could be
magnetospheric in origin, we argue that damping between the neutron
star crust and interior liquid is a more likely explanation. Under
this interpretation, our results provide the first evidence
independent from pulsar spin glitches of differential rotation in
neutron stars.

While this paper focuses on identification of signatures of the
underlying physical processes responsible for pulsar timing
fluctuations, the mathematical characterization of timing noise also
plays a crucial role in efforts to detect gravitational waves in
pulsar timing arrays such as the North American Nanohertz Observatory for Gravitational Waves {\sl NANOGrav} (http://www.nanograv.org), the Parkes Pulsar Timing Array {\sl PPTA} ( http://www.atnf.csiro.au/research/pulsar/ppta/), and the European Pulsar Timing Array {\sl
EPTA} ( http://www.epta.eu.org). In particular, if the underlying power spectrum of timing noise
is known, the signal-to-noise ratio for the gravitational wave
background could be substantially improved
\citep{vanhaasteren_etal09}.

In \S 2 we describe the data that we analyzed. In \S 3 we describe our
analysis methods. In \S 4 we present results of the analysis for five
radio pulsars. In \S 5 we discuss our results.  The details of the analysis methods, examples of applications to simulated data sets, and tests of robustness are presented in the Appendix.

\section{Data}

The Green Bank data were collected using a 25-m radio telescope at the
National Radio Astronomy Observatory, Green Bank, West Virginia. This
telescope monitored the pulsars on a near daily basis from 1989
through 1999.  Observations followed a fixed daily schedule (adjusted
a few times over the course of the program).  Integration time
depended on source strength, but was typically of order 40 minutes,
and was divided into several subintegrations.  For each of two linear
polarizations, a filter bank produced total-power values for each of
sixteen 1-MHz spectral channels across a band centered at 610 MHz.
(The channels were not completely contiguous due to radio frequency
interference considerations.).  The two polarizations were balanced
(using measured system noise), summed, and folded modulo the pulse
period using the Princeton Mark 3 data acquisition system
\citep{stinebring_etal92}. 
This procedure produced a single pulse profile for each spectral
channel in each subintegration.  Off-line, the folded data profiles
were cross-correlated with a high precision standard template to
produce pulse times of arrival (TOAs).  The TOAs from all channels and
all subintegrations of any given pulsar on any given day were averaged
to produce a single effective daily TOA, which was used for the
present analysis. These TOAs were then analyzed with the {\sc tempo}2
software package \citep{hobbs_etal06} to obtain the timing residuals, using the JPL DE405 ephemeris.  In each case, fits were made for position, proper motion, rotational frequency, and rotational frequency derivative.

The Jodrell Bank data were collected using the 13-m Lovell Radio
Telescope with nearly daily observations from 1991 to 2008.
Observations were made at 610 MHz, with a 4 MHz bandwith.  30-60 minutes of
observations were made for each pulsar, depending on the flux density
of each object, and then divided into 1-3 minute subintegrations.  The
subintegrations were averaged to produce a single profile, which was
convolved in the time domain with the corresponding pulse template to
produce TOAs \citep{hobbs_etal04}. These TOAs were then analyzed with
the {\sc tempo}2 software package \citep{hobbs_etal06} to obtain the timing
residuals, using the JPL DE200 ephemeris.

\section{Analysis Methods}

\begin{figure}

{\includegraphics[scale=.8]{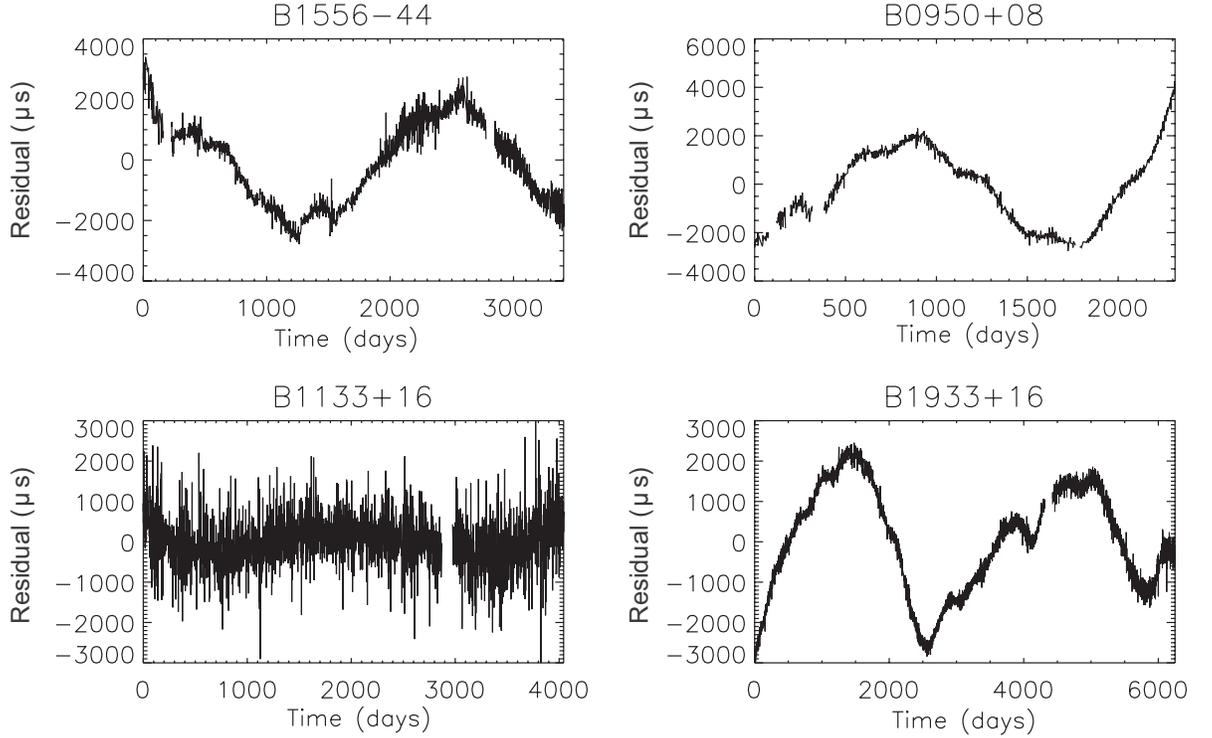}}
\centering
\caption{Timing residuals in microseconds for four pulsars, 
showing typical long-period wander. The data are from
the Jodrell Bank and Green Bank data archives (see Table 1). The residuals were obtained using the
{\sc tempo}2 software package (Hobbs et al. 2006).}
\label{fivefigs}
\end{figure}

Timing noise appears as stochastic wander of the pulse arrival time,
with respect to a deterministic spin-down model, over all timescales;
examples are shown in Fig. \ref{fivefigs}. In most cases, the dominant
effect is that of low-frequency variation over years. To study
possible correlations over much shorter timescales, we work in the
time domain and first subtract the slow wander of the timing residuals through 
application of a high-pass filter.
We then perform two types of correlation
analyses on these filtered (that is, ``whitened'') data.

\subsection{High-Pass Filtering}

We divide the time series into contiguous non-overlapping intervals of
width $W$ and calculate the average value of the residuals in each
interval. Using unweighted least-squares fitting, we fit the
average values with a cubic spline, and subtract the spline from the
original time series to obtain the whitened residuals.
More details and examples of this filtering method are given in the
Appendix. For most data sets, a broad range in choices of the filter
width $W$ effectively removes long-period wander without introducing
spurious correlations.

\subsection{Discrete Correlation Function}

The data we analyze are unevenly sampled, and often contain large
gaps. It is desirable to measure autocorrelations without having to
bin the data, which would restrict our analysis to a timescale no
shorter than the longest gap in the data and would entail a severe loss in
time resolution. Unevenly-sampled data is readily handled with the
{\em discrete correlation function} ${\rm DCF}(\tau)$ \citep{ek88}, which we
calculate as follows. For a set of arrival time residuals $\delta t_i$
measured at times $t_i$, we construct the matrix
\begin{equation}
{\rm DCF}_{ij} = \frac{(\delta t_i-\overline{\delta t})(\delta
t_j-\overline{\delta t})} {\sigma^2},
\label{matrix}
\end{equation}
where $\overline{\delta t}$ is the mean of the data set and $\sigma$
is its standard deviation (the choice of normalization is
arbitrary). This matrix is calculated for all possible pairs $(\delta
t_i,\delta t_j)$, each of which is associated with the time
difference $\Delta t_{ij}\equiv t_j - t_i$ between the $i$-th and
$j$-th measurements. (Here and in the following, $\delta t$ will
denote residuals, and $\Delta t$ will denote differences in
measurement times). Suppose that there are $M$ pairs that satisfy
the condition
\begin{equation}
\tau - \Delta \tau/2 \le \Delta t_{ij} < \tau + \Delta \tau/2, 
\label{tarray}
\end{equation}
where $\Delta\tau$ is the width of the sampling window. The discrete
correlation function is the time average of eq. (\ref{matrix}) for
the pairs that satisfy eq. (\ref{tarray}), that is, 
\begin{equation}
{\rm DCF}(\tau) = \frac{1}{M}\sum {\rm DCF}_{ij}.
\end{equation}
Points that do not fall in the sampling window do not contribute to
${\rm DCF}(\tau)$. The window size is chosen to maximize resolution
without loss of statistical significance; in practice, $\Delta\tau$
can be taken to be almost as small as the average sampling time. This
procedure uses every data point, without any significant penalty in
resolution due to occasional large gaps in the data. The values of
$\tau$ that are used are binned in units of $\Delta\tau$.
The standard uncertainty in the ${\rm DCF}(\tau)$ is
\citep{ek88}
\begin{equation}
\sigma_{\rm DCF}(\tau) = \frac{1}{M-1}\left\{\sum \left [{\rm DCF}_{ij}-
{\rm DCF}(\tau)\right ]^2\right\}^{1/2}. 
\label{dcfbar}
\end{equation}

\subsubsection{Lagged Dispersion}

As another statistic with which to measure correlations, we introduce
the {\em lagged distribution function}, $\rm{LDF}\{\delta t,\tau\}$, the
distribution of fluctuation {\em differences} separated in time by a
lag $\tau$. Correlations in the data can be studied with the {\em
lagged dispersion} ${\rm LD}(\tau)$, which we define as the dispersion of the lagged
distribution function.  The LD has the following properties for a data
set that is correlated over a timescale $\tau_c$ (and without a
resonance). Because the time series will resemble itself to some
extent upon time translation by times $\tau<\tau_c$, the $\rm{LDF}\{\delta
t,\tau\}$ will be relatively narrow, and hence $LD(\tau)$ will be
relatively small. For $\tau>\tau_c$, the dispersion is larger because
the data around time $t+\tau$ are uncorrelated with the data around
time $t$; in this case $\rm{LDF}\{\delta t,\tau\}$ is broader than at low
lag, and ${\rm LD}(\tau)$ asymptotes to some maximum value as $\tau$ is
increased. If the data series is simply noise, there will be no
statistically-significant variations in ${\rm LD}(\tau)$ with $\tau$, since
the data contain no timescale. A general increase of ${\rm LD}(\tau)$ up to
a lag $\tau\simeq\tau_c$ indicates that the series is correlated over
a timescale $\sim\tau_c$. Oscillations in the data appear also as
oscillations in the LD. 

To handle uneven sampling, we construct ${\rm LD}(\tau)$ in a similar 
way as the $DCF(\tau)$: 
\begin{equation}
{\rm LD}(\tau)\equiv\frac{1}{M-1}\sum (\delta t_{ij} -
\overline{\delta t})^2, 
\label{disp}
\end{equation}
where $\delta t_{ij}\equiv \delta t_i-\delta t_j$. 
As before, the product is over the set of $M$ elements that
satisfies eq. (\ref{tarray}), in which each element of the set is
associated with the lag value $\tau$. The mean $\overline{\delta t}$ in
eq. (\ref{disp}) is defined as
\begin{equation}
\overline{\delta t}\equiv\frac{1}{M}\sum \delta t_{ij}.
\end{equation}
The LD is mathematically similar to the structure function used in
studies of the turbulent interstellar plasma ({\sl e.g.},
\citealt{rickett90,you_etal07}).

The standard uncertainty in ${\rm LD}(\tau)$ is
\begin{equation}
\sigma_{\rm LD}(\tau) = \frac{1}{M-1}\left [\sum (\delta t_{ij}-
\overline{\delta t})^2 - \rm LD(\tau) \right ]^{1/2}.
\label{error}
\end{equation}
 
In the Appendix, we illustrate the usefulness of the DCF and the LD, in
combination with high-pass filtering, to identify short-timescale
correlations that are not readily identified with Fourier techniques.
We find that high-pass filtering of the time series followed by
calculation of the DCF or LD is a robust method for identifying an intrinsic
relaxation timescale $\tau_c$, provided the following conditions are
satisfied:
\begin{equation}
   \Delta t_{\rm samp} < \tau_c < \tau_{\rm wander},
\end{equation}
where $\Delta t_{\rm samp}$ is the mean sampling interval and
$\tau_{\rm wander}$ is the shortest timescale of the wander.  If the first
inequality is not satisfied, then the time resolution of the data is
not sufficient to resolve the correlation timescale. If the second
condition is not met, then the correlation cannot be disentangled from
the wander. In practice, we must also require
\begin{equation}
\tau_{\rm corr}<W<\tau_{\rm wander}, 
\end{equation}
to ensure that the filter removes the wander but not the
correlation. In the extreme case of $W\sim \tau_{\rm samp}$, the
filtering method introduces spurious {\rm anti}-correlations,
indicating that the choice of $W$ is too small. 

\section{Correlation Analyses}

\begin{table}
\centering
\begin{tabular}{cccccc}
\hline
 Name & Observatory &  Period (s) & $\tau_{\rm age}$ (Myr)  & Data Span (d) & No. of Observations  \\
\hline

$B0136+57$ & GB & 0.27 & 0.4 & 3558 & 2685 \\ 
$B0329+54$ & JB & 0.72 & 5 & 4313 & 3884 \\
$B0355+54$ & JB & 0.16 & 0.6 & 3376 & 855 \\
$B0525+21$ & GB & 3.75 & 1 & 3387 & 1346 \\ 
$B0628-28$ & GB & 1.24 & 3 & 3177 & 1593 \\ 
$B0736-40$ & GB & 0.37 & 4 & 2312 & 1056 \\ 
$B0740-28$ & JB & 0.17 & 0.2 & 1820 & 1631 \\ 
$B0818-13$ & GB & 1.24 & 9 & 3402 & 1680 \\ 
$B0823+26$ & GB & 0.53 & 5 & 2113 & 1025 \\
$B0835-41$ & GB & 0.75 & 3 & 3561 & 1948 \\
$B0950+08$ & GB & 0.25 & 20 & 2312 & 1223 \\
$B1133+16$ & JB & 1.19 & 5 & 4043 & 3956 \\
$B1237+25$ & GB & 1.38 & 20 & 2267 & 923 \\
$B1508+55$ & GB & 0.74 & 2 & 3561 & 1923 \\
$B1556-44$ & GB & 0.26 & 4 & 3405 & 2498 \\
$B1641-45$ & GB & 0.46 & 0.4 & 3556 & 2071 \\
\hline

$B1642-03$ & JB & 0.39 & 3 & 6248 & 5555 \\ 
$ B1749-28$ & GB & 0.56 & 1 & 3554 & 2205 \\ 
$ B1818-04$ & GB & 0.60 & 2 & 3558 & 1877 \\ 
$ B1822-09$ & GB & 0.77 & 0.2 & 1091 & 476 \\ 
$ B1831-03$ & GB & 0.69 & 0.3 & 3396 & 1711 \\ 
$ B1859+03$ & GB & 0.66 & 1 & 3176 & 1762 \\
$ B1911-04$ & GB & 0.83 & 3 & 2310 & 1224 \\ 
$ B1919+21$ & GB & 1.34 & 20 & 3548 & 2567 \\ 
$B1929+10$ & JB & 0.23 & 3 & 4267 & 3800 \\
$ B1933+16$  & JB & 0.36 & 0.9 & 6249 & 5377 \\
$B1946+35$ & GB & 0.72 & 2 & 2171 & 1163 \\
$B2016+28$ & GB & 0.56 & 60 & 1885 & 951 \\
$B2020+28$  & GB & 0.34 & 3 & 1953 & 1238 \\
$B2045-16$ & GB & 1.96 & 3 & 1705 & 699 \\
$B2111+46$ & GB & 1.02 & 20 & 3544 & 1986 \\
$B2217+47$ & GB & 0.54 & 3 & 3556 & 1971 \\

\hline
\end{tabular}
\caption{The pulsars to which we have applied the analysis techniques
of this paper.  JB - Jodrell Bank timing data, from the 13m dish with an observing frequency of 610 MHz, with a 4 MHz bandwidth, using 30-60 min observing sessions.  GB -
Green Bank timing data, from telescope 85-3 (25m diameter), with an observing frequency of 610 MHz and 16 MHz bandwith, with typical observation sessions of $\sim$ 40 min. 
The pulsar parameters given are from ATNF Pulsar Catalog
(http://www.atnf.csiro.au/people/pulsar/psrcat/). }
\label{pulsars}  
\end{table}

We have applied the methods of \S 3 to the 32 radio pulsars of Table
\ref{pulsars}. We apply a high-pass filter to the residuals, and
then calculate the ${\rm DCF}$ and the ${\rm LD}$. In most cases, we
see no statistically significant correlations after the data are
whitened. Here we present results for PSRs B1133+16, B1933+16,
B0525+21, B1556-44, B0950+08, as particularly interesting examples
that show correlated structure in their time series.

\begin{figure}
\centering
{\includegraphics[scale=.80]{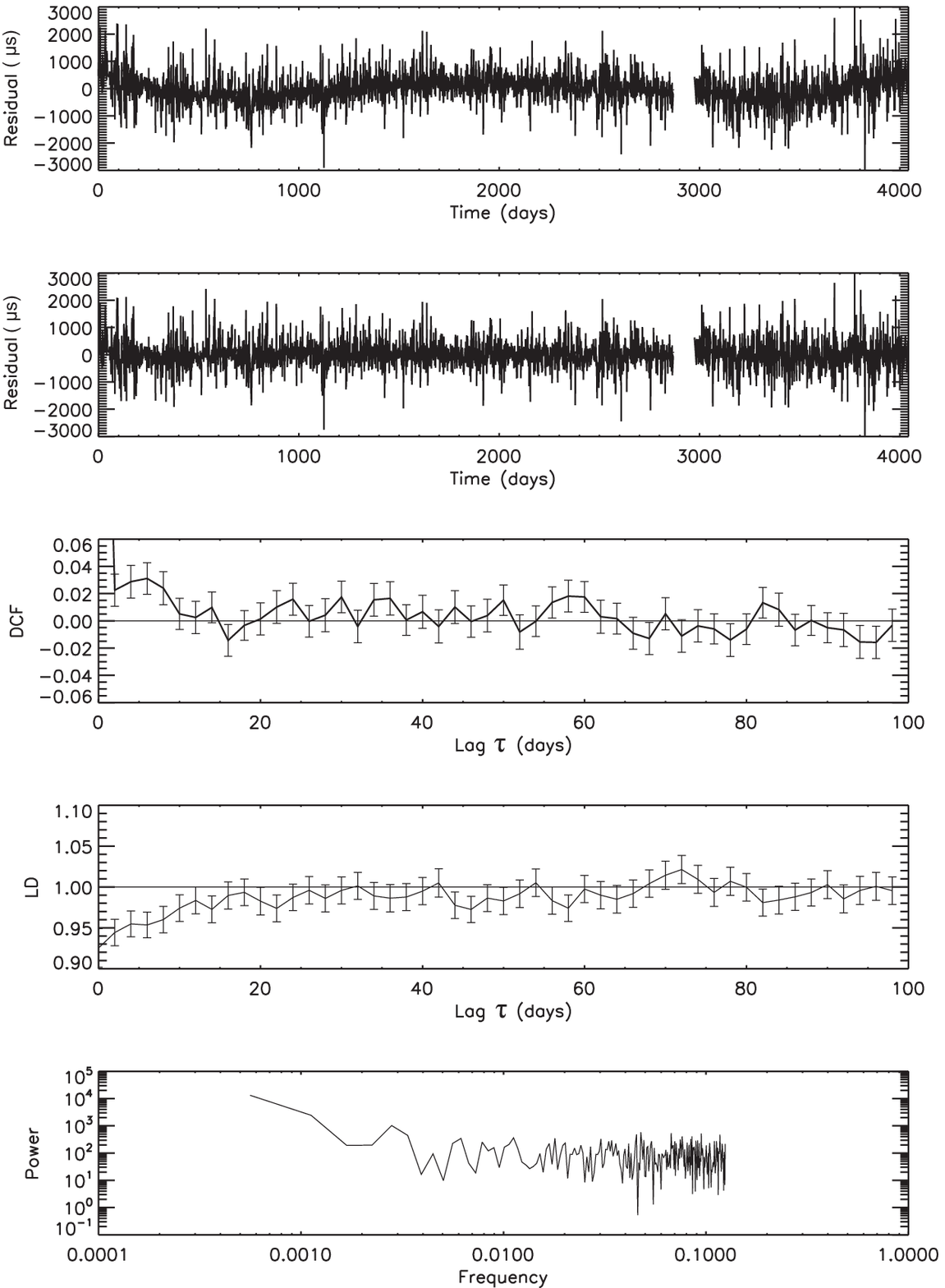}}
\caption{Timing fluctuations in PSR B1133+16, in microseconds. {\sl Second panel} - 
timing fluctuations after subtraction of the
long-term wander evident in the top panel, using a filter width
$W=400$ days.  The uncertainties are comparable in magnitude to the
fluctuations, and are not shown for clarity. The gaps in the data sets
are on account of equipment upgrades; our analysis techniques enable
us to use data on both sides of the gaps. {\sl Third panel} - DCF of
the whitened residuals.  {\sl Fourth panel} - LD of the whitened
residuals.  {\sl Bottom panel} - power spectrum of the 1700 day span of unwhitened residuals (see text). }
\label{resids1}
\end{figure}

{\bf{PSRs B1133+16 and B1933+16.}}  These two pulsars have been
monitored almost daily with the 12.8 m telescope at Jodrell Bank,
usually at 610 MHz. Timing residuals for PSR B1133+16 over about 4000
days, and PSR B1933+16 over about 6000 days, are shown in
Fig. \ref{resids1}.  We begin with these pulsars because they show
particularly interesting correlations in their times series.

\begin{figure}
\centering
{\includegraphics[scale=.80]{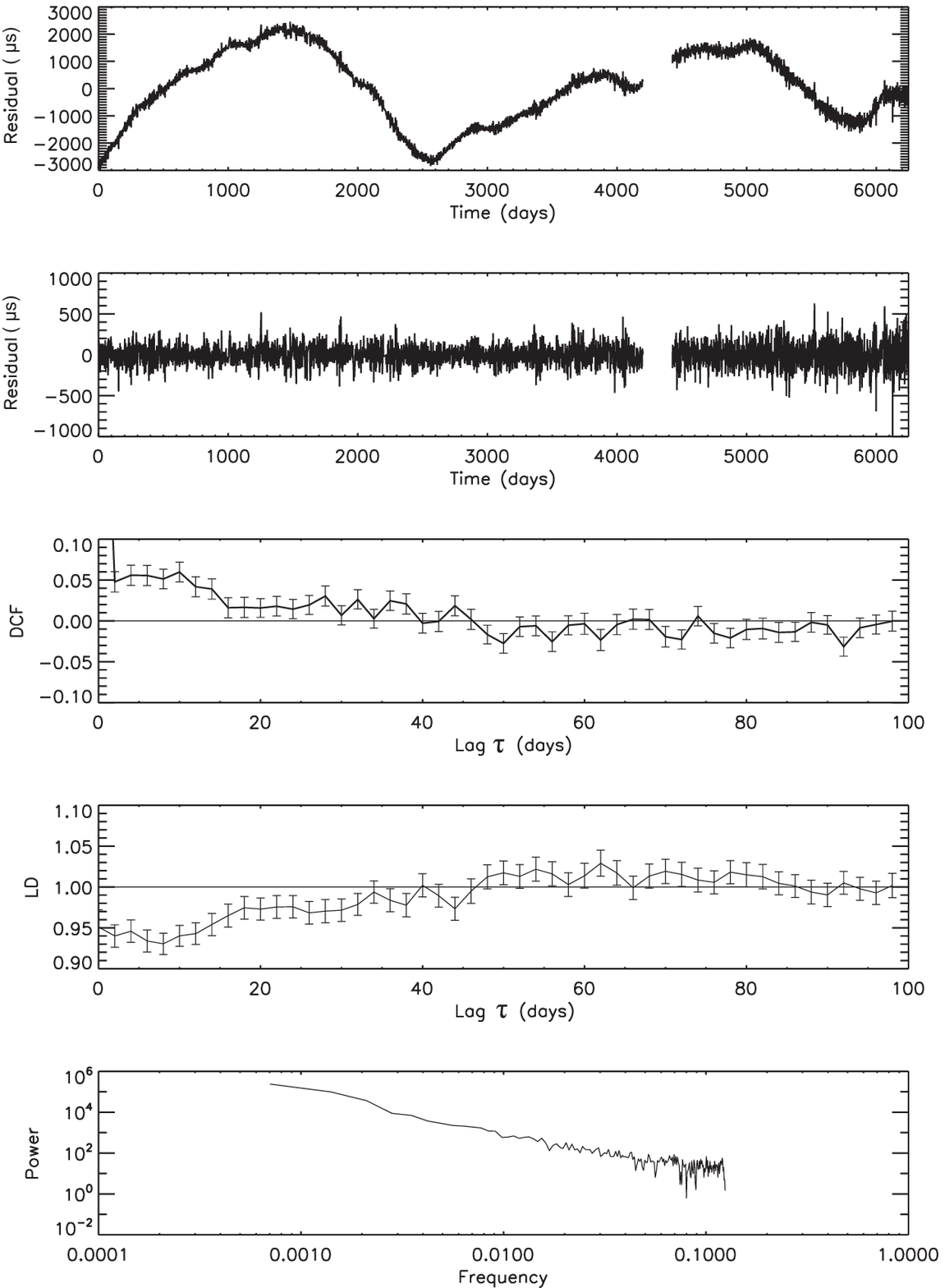}}
\caption{Same as Fig. \ref{resids1}, for PSR B1933+16. 
The filter width is $W=120$ d, for which the correlations over
timescales of $\sim 20$ d are most clearly seen.  The power spectrum in the bottom panel is for the 1400 day span of unwhitened residuals (see text).} 
\label{resids2}
\end{figure}

In Figs. \ref{resids1} and \ref{resids2} (3rd panel) we show the
DCFs for B1133+16 and B1933+16, after whitening with filter widths $W$
of 400 d and 120 d, respectively. Uncertainties were obtained
from eq. (\ref{dcfbar}). (All uncertainties in this paper are
1-$\sigma$). The DCF for B1133+16 shows highly significant,
non-periodic correlations over a timescale $\tau_c\sim 10$ d, with the
distinct structure of a relaxation process (see Appendix). B1933+16 shows similar
correlations, over a timescale $\tau_c\sim 20$ d.  From the calculated
uncertainties, shown in the figures, the confidence level of the
detected correlations is high.  We estimate the correlation
significance to be 
\begin{equation}
   S = \prod_i \frac{{\rm DCF} (\tau_i)} { \sigma (\tau_i)},
\end{equation}
where the product is over all values of $\tau_i < \tau_c$, the
inferred correlation timescale. The quantity $S$ should give an
approximate measure of the significance of the result in units of the
(typical) statistical uncertainty of each data point. For B1133+16,
the correlation over $\tau_c=10$ is seen over the first four bins of
width $\Delta\tau=2$ d, giving $S=24.5$. For B1933+16, the correlation
over $\tau_c=20$ is seen over the first nines bins of width
$\Delta\tau=2$ d, giving $S=31.2$. These values of $S$ indicate very
high statistical significance. As a check, we obtain a lower limit on
the significance with a bootstrapping method. We create an ensemble of
data sets in which the time-ordering of the whitened residuals is
randomly shuffled. The DCF is calculated for the shuffled data, with
$\Delta=2\tau$ d for both pulsars.  We calculate the significance
$S_{\rm shuffled}$ for $0\le\tau\le\tau_c$ (the first four bins for
PSR B1133+16, and the first nine bins for PSR B1933+16). We then ask:
what fraction of shuffled sets give $S_{\rm shuffled}>S$?  Under the
null hypothesis of uncorrelated data, $S_{\rm shuffled}$ is comparable
to $S$, since $S$ represents one particular realization. In $10^4$
shufflings for each data set, the null result was never realized.  We
therefore estimate the statistical significance of the detected
correlations to exceed $1-10^{-4}$, consistent with the values of $S$
obtained above. This estimation method has the advantage that nothing
is assumed about the underlying statistics of the data, rather, the
data themselves are used to evaluate the likelihood of the null
hypothesis.

In Figs. \ref{resids1} and \ref{resids2}, we show ${\rm LD}(\tau)$ for
PSRs B1133+16 and B1933+16. The value of LD for uncorrelated data is
calculated by averaging many shufflings of the data; this value is
used to normalize ${\rm LD}(\tau)$ shown in the figures (see Appendix
A3). For B1133+16, we find that ${\rm LD}(\tau)$ is significantly
lower than what we expect from the null hypothesis out to $\tau\simeq
10$ d, in support of the DCF result.  B1933+16 is very similar, but
indicates a relaxation timescale $\tau_c\simeq 30$ d, consistent with
the DCF.

The power spectra of the unwhitened data do not show the correlations that
our time-domain analysis has revealed.  Gaps in observations of 4-5
days are frequent for B1133+16, and several gaps of 8-16 days are
present.  To obtain a power spectrum without interpolating data, we
have chosen a span of data approximately 1700 days in length, with
gaps no longer than four days.  The power spectrum of the timing
residuals are shown in Fig. \ref{resids1} (bottom panel), using
uniformly weighted bins of width four days. The spectrum is
generally white, with some excess power at low frequencies.  No high
frequency features or spectral breaks are evident.  For PSR B1933+16,
we select a segment of data spanning 1400 days, also placed in bins of
width four days, to calculate the power spectrum
shown in Fig. \ref{resids2} (bottom panel).  The red spectrum
corresponds to the low-frequency wander visible in the timing
residuals (Fig. \ref{resids2}).  The Fourier transform spreads the
power over all frequencies, obscuring correlations that are more
readily identified in the time domain.

The two complementary and independent statistics we have used here,
${\rm DCF}(\tau)$ and ${\rm LD}(\tau)$, measure different properties of the
data. The DCF measures the extent to which the data set compares to
itself upon translation in time, while the LD measures the extent to
which the distribution of fluctuation differences widens as the data
decorrelate for larger lags. These two different statistics give
consistent results for PSRs B1133+16 and B1933+16.  Several tests of
the robustness of these results are given in the Appendix. In
particular, the whitening process cannot introduce spurious
correlations at timescales shorter than the window width. The filter
width of $W=400$ d used for the analysis of PSR B1133+16 is a factor
of $\sim 40$ larger than the derived correlation time.  The filter
width of $W=120$ d used for the analysis of PSR B1933+16 is a factor
of $\sim 6$ larger than the correlation time.

\begin{figure}
\includegraphics[scale=1.0]{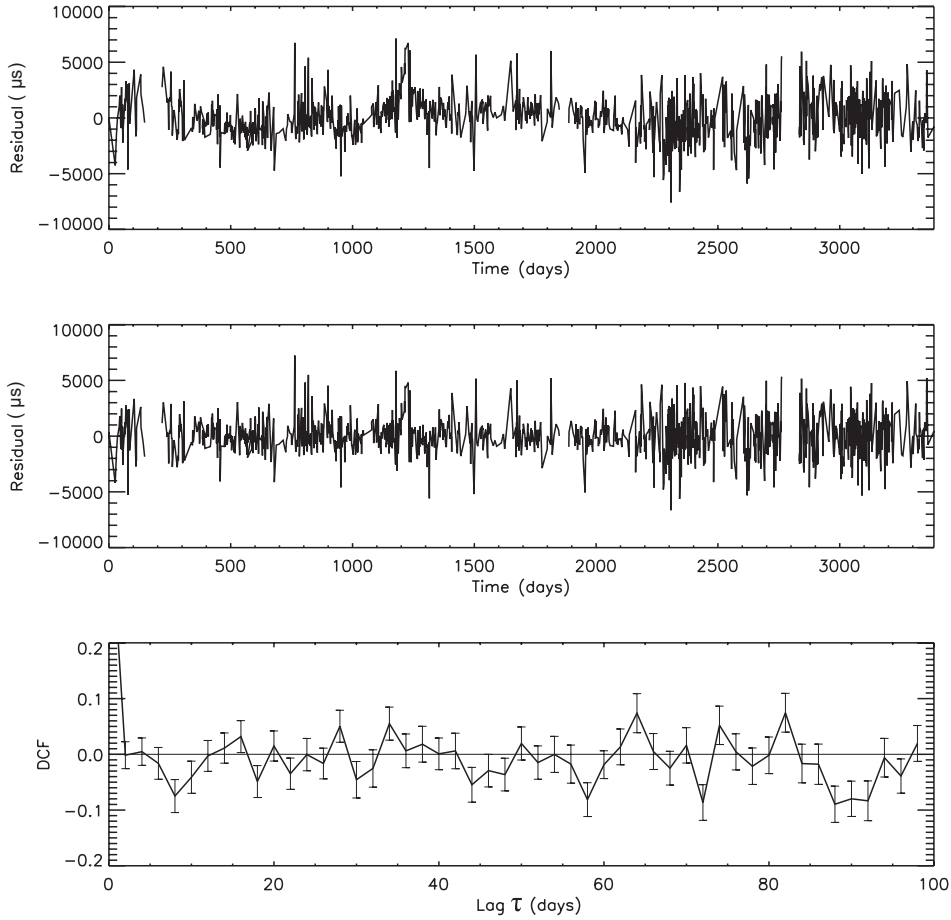}
\caption{{\sl Top panel} - Timing residuals for PSR B0525+21, in
microseconds.  {\sl Middle panel} - residuals after whitening with
W=100 days.  {\sl Bottom panel} - The DCF of the whitened data,
showing variability over $\sim 10$ d. }
\label{0525two}
\end{figure}

{\bf PSR B0525+21.}  Timing residuals are shown
in Fig. \ref{0525two} (top panel).  The wander in the timing noise
occurs over timescales $\gtrsim 300$ days.  Upon whitening with a
filter width $W=100$ days, the DCF (bottom panel) shows 
variability over timescales of $\sim 10$ d, but no evidence of
relaxation. Relaxation response, if present, could be overwhelmed by
the variability.

\begin{figure}
\centering
\includegraphics[scale=1.0]{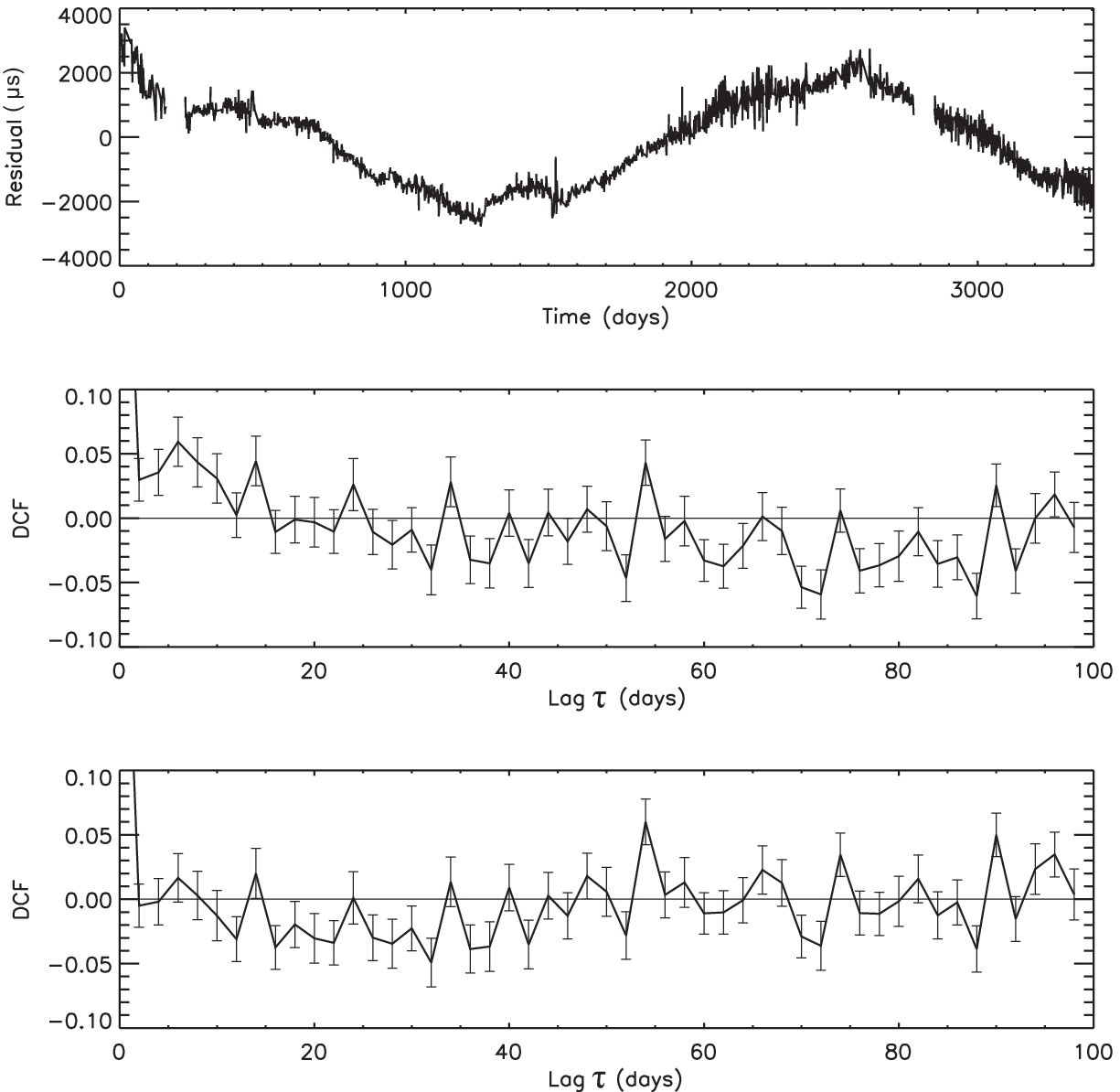}
\caption{{\sl Top panel} - Timing residuals for PSR B1556-44.  
{\sl Middle panel} - The DCF after whitening with a filter width
$W=100$ d.  Variability over $\gap 10$ d is evident.  {\sl Bottom panel} -
The DCF with $W=50$ days.  More wander has been
removed, but variability remains.}
\label{1556all}
\end{figure}

{\bf PSR B1556-44.} Timing
residuals are shown in Fig.
\ref{1556all} (top panel).  DCFs for this pulsar are shown in Fig.
\ref{1556all} for $W=100$ (middle panel) and 50 days (bottom panel).  For 
$W=100$ days, the DCF shows variability over $\sim 10$
d. With $W=50$ days, more wander is removed, but variability over
timescales $\gap 10$ d is still evident (bottom panel). As for PSR
B0525+21, any possible signature of relaxation response might be
overwhelmed by the variability. 

\begin{figure}
\includegraphics[scale=.8]{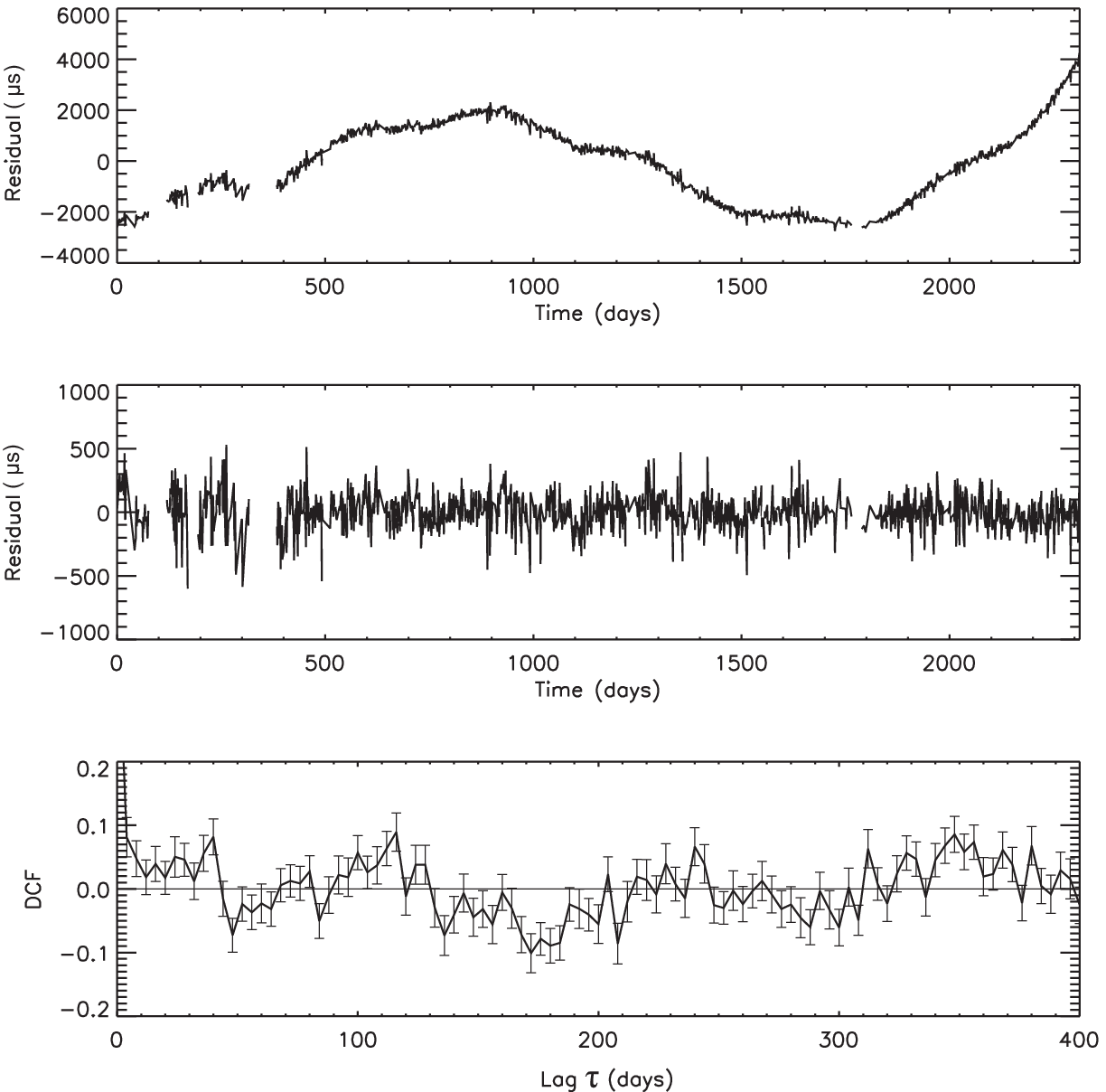}
\caption{
{\sl Top panel} - Timing residuals for PSR B0950+08, in microseconds.
{\sl Middle panel} - The residuals after whitening with a filter width
$W=100$ d. {\sl Bottom panel} - The DCF of the whitened
residuals, showing variability over timescales comparable
to observed pulse mode changes \citep{shab04}.}
\label{0950res}
\end{figure}

{\bf PSR B0950+08.} Timing residuals are shown in Fig. \ref{0950res}.
Upon whitening the data, the DCF of this pulsar shows both
correlations and anti-correlations (Fig. \ref{0950res}, bottom panel), 
associated with quasi-periodic variability over a timescale of $\lap
100$ d.  The pulse profile of PSR B0950+08 changes modes over
timescales comparable to those identified in the DCF \citep{shab04};
the variability in the DCF could be directly related to mode changing.

\section{ Discussion} 

We find strong evidence for correlation of timing residuals over
timescales less than $\sim 10$ d in PSR B1133+16, and $\sim 20$ d in
PSR B1933+16, indicative of relaxation response in the neutron star
system of crust, core, and magnetosphere. While it is possible that
the correlations are created by variable torques in the magnetosphere,
we consider this possibility to be unlikely. The basic timescale in the
magnetosphere is the light travel time across the light cylinder,
which is comparable to the spin period. For the observed correlations
to be magnetospheric in origin, the magnetosphere must have a
relaxation time of weeks. Such a long relaxation time is difficult to
explain on physical grounds. \citet{lyne10} have shown that many stars
exhibit changes in spin-down rate in association with pulse-shape
changes, clear evidence that the magnetosphere is dynamic, but these
changes always occur much more quickly than $\sim 10$ d.  PSR B1133+16
and PSR B1933+16 show no evidence for such modes changes, though PSR
1133+16 does show nulling. Mode changes would show both correlations
and anti-correlations, unlike B1133+16 and B1933+16, which show only
correlations.

The correlations cannot be due to time variability of the
interstellar medium.  The delay in the arrival time of a pulse of
frequency $\nu_{\rm MHz}$ due to dispersion in the interstellar medium is
\citep{lyne_gs05}
\begin{equation}
 \Delta t_{ISM} = 4.15 \times 10^{6} \,\rm{MHz}^2 \,\rm{pc}^{-1} 
{\rm cm}^3\,\rm{ms} \,\times \nu_{MHz}^{-2} \times DM,
\end{equation}
where DM is the dispersion measure (cm$^{-3}$ pc).  The dispersion
measure is calculated during multi-wavelength observations for
B1133+16 and B1933+16, and used to subtract the delay as part of the
timing solution.  The time derivative of the dispersion measure has
been measured from long-term monitoring: $d(DM)/dt\sim 8 \times
10^{-4} \rm{cm}^{-3} \,\rm{pc} \,\rm{yr}^{-1}$ for B1133+16, and
$\sim2.3\times10^{-3} \,\rm{cm}^{-3} \,\rm{pc} \, \rm{yr}^{-1}$ for
B1933+16 \citep{hobbs_etal04}.  This variability of the electron density
produces small fluctuations in the arrival time delay, on the order of
$10 \, \rm{\mu} \rm{s} \, \rm{yr}^{-1}$. To estimate the maximum
effect of variations of the interstellar medium, we superimposed a
sine wave of amplitude 10 $\mu$s with period 10 d on our whitened
data. No detectable signal is introduced into either the DCF or the
LD.

An alternative explanation is that the correlations found for PSRs
B1133+16 and B1933+16 are due to damping between the neutron star
crust and interior liquid as the system is excited, presumably often,
away from a state of rotational equilibrium that is never reached. In
this picture, our results provide the first evidence independent from
pulsar spin glitches of differential rotation in neutron stars. In
support of this interpretation, we note that the measured correlation
times are comparable to the post-glitch recovery timescales measured
in many pulsars ({\sl e.g.}, \citealt{lss00}).

The methods introduced here are also generally useful for identifying
quasi-periodic processes that occur over timescales of days. PSR
B0950+08 is a particularly clear example, showing alternating
correlations and anti-correlations in the DCF. This star exhibits mode
changes over a similar timescale \citep{shab04}, and so the DCF
appears to be showing magnetospheric effects. Though not as
strong as PSR B0950+08, PSRs B0525+21 and B1556-44 also show evidence
for variability.

For most of the pulsars in Table \ref{pulsars}, we find no evidence of
relaxation behavior. In order to resolve correlation timescales of
$\sim 10$ d, nearly daily sampling is required.  Most of the pulsars
that we have analyzed do not have the high sampling rates of PSRs
B1133+16 and B1933+16. These pulsars with lower sampling rates might
have short-timescale correlations that are not detectable yet with the
methods used here. For data which contain relaxation and periodicity
over similar timescales, it is not possible to remove the wander
without removing any non-periodic correlations that may exist.

The analysis methods applied are suitable for seeking time correlated
structure in any noisy time series (see, {\sl e.g.},
\citealt{fukumura10} for a recent application of autocorrelation
methods to the analysis of astronomical data).  For pulsar timing noise in
particular, these techniques show promise for distinguishing effects
that might be magnetospheric in origin, such as mode switching, from
those effects that are directly related to rotational response of the
stellar interior. Mode switching, for example, is now known to
be related to sudden changes in magnetospheric torque in many pulsars
\citep{lyne10}. This effect can be seen as quasi-periodic behavior of the
DCF, as we have shown with PSR B0950+08. By contrast relaxation
response, as seen in PSRs B1133+16 and B1933+16, is more
naturally attributed to internal dynamical degrees of freedom.  An
in-depth study of relaxation response in the context of a physical
model is in progress. The techniques developed in this paper may also
be used to detect precession and oscillation modes in pulsar timing
data.

\section*{Acknowledgements}

We are grateful to Michael Kramer and Andrew Lyne for preparing the
data from the Jodrell Bank archive used in this paper, and for
valuable discussions and comments on an early version of the
manuscript.  S. P. and B. L. thank the Department of Physics,
University of Pisa, and the INFN -- Sezione di Pisa for support of this
project. B. L. acknowledges support from U. S. NSF grant AST-0406832.
S.P. thanks the Montana Space Grant Consortium for support during the
time of this work.

The National Radio Astronomy Observatory is a facility of the National
Science Foundation operated under cooperative agreement by Associated
Universities, Inc.  Operation of the 26~m telescope at Green Bank was
supported by the US Naval Observatory.  D. J. N. thanks NSF for
support via grant AST-0647820 and previous awards. 

We thank B. Coles, J. Cordes, G. Hobbs, V. Kaspi, Y. Levin, and
D. Werthimer for valuable discussions and assistance.

\bibliographystyle{mn2e}
\bibliography{References2}

 \appendix
 
 \section{Examples and Tests}

In this Appendix we describe the analysis techniques used in this
paper in more detail and present many examples and tests of
robustness. Synthetic data sets are generated to simulate the
properties of real radio pulsar timing data. We also discuss a simple
physical model for random spindown torques that captures the essential
features of the detected correlations.

 \subsection{Whitening with a High-Pass Filter}
 
  \begin{figure}
\centering
{\includegraphics[scale=.8]{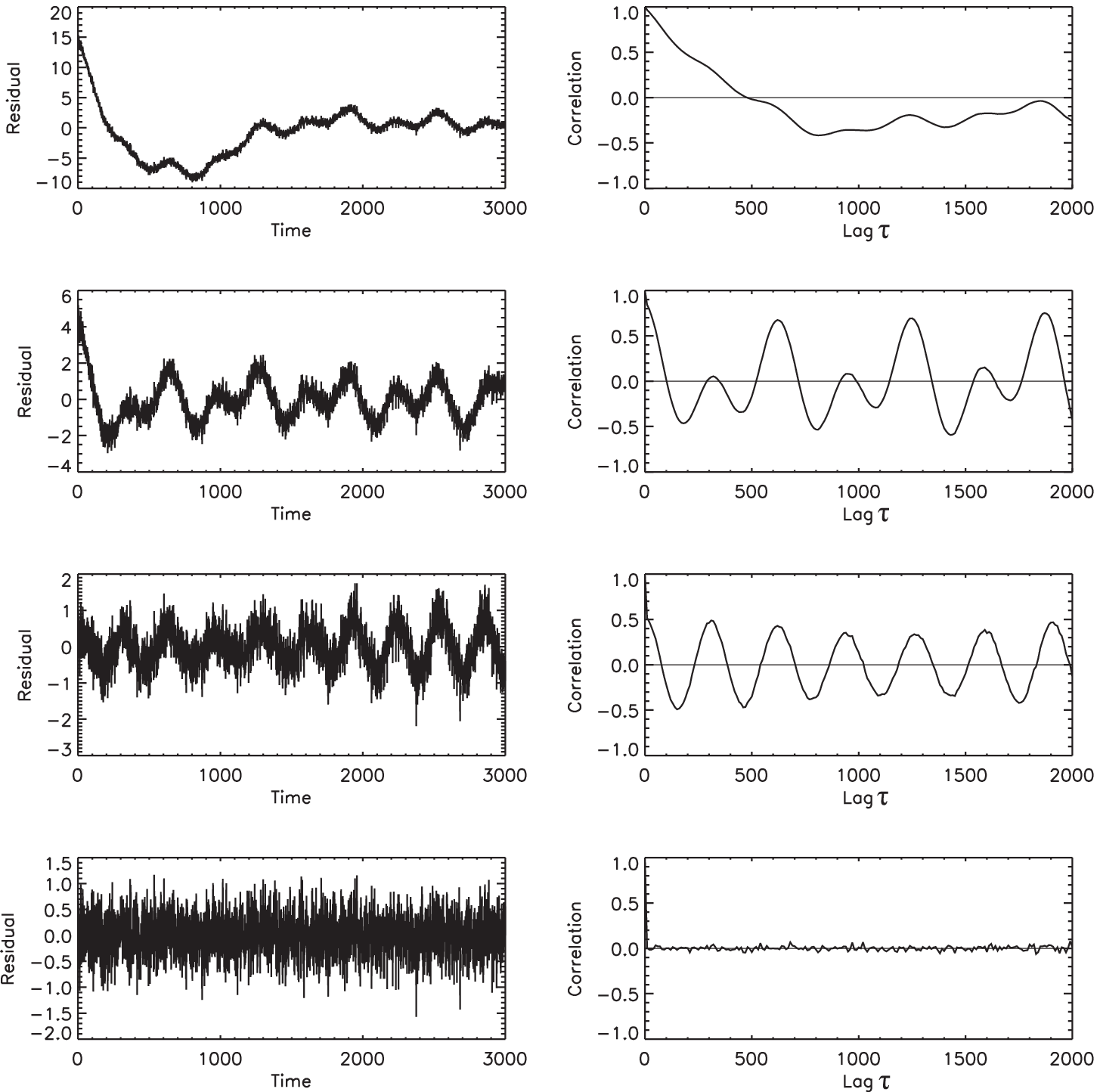}}
\caption{{\sl Top panel} - Simulated residuals (arbitrary units)
consisting of 10 sine waves, an arbitrary 5th order polynomial, and
Gaussian noise, and the corresponding ACF. {\sl Second panel} -
Whitened data with a filter width $W=300$.  {\sl Third panel}- Further
whitening with filter width $W=150$.  {\sl Bottom panel}- Further
whitening with filter width $W=75$.  }
\label{synthetic}
\end{figure}
 
 To illustrate the effects of the high-pass filter that we use, we
 construct simulated timing residuals $\delta t(t)$ consisting of
 periodic functions, quasi-periodic wander and gaussian noise,
\begin{equation}
 \delta t(t) = \sum_i \left ( \frac{A_i} {\omega_i} \sin(\omega_i t + \phi_i) \right ) + P(t) + N(t), 
\end{equation} 
where $A_i$, $\omega_i$, and $\phi_i$ are the amplitude, frequency,
and phase, respectively, $P(t)$ is an arbitrary polynomial function,
and $N(t)$ is gaussian noise.  $A_i$ and $\phi_i$ are randomly
generated values.  The sine waves are weighted by the frequency
$\omega_i$ to produce a red power spectrum, similar to that of typical
pulsar data.  An example of a simulated time series is shown in
Fig. \ref{synthetic} (top panel), using a sum of 10 sine waves.
For this example, we choose frequencies $\omega_i=\omega_1/i$, where
$i$ is an integer, and a weighting of $\omega_i^{-1}$. The maximum
frequency is $\omega_1 = 0.32$ (period=19.6), and and the minimum
frequency is $\omega_{10} = 6.25 \times 10^{-4}$ (period=$10^4$).

We now whiten the synthetic data as follows. 
We divide the time series into contiguous non-overlapping intervals of
width $W$ and calculate the average value of the residuals in each
interval.  Using unweighted least-squares fitting, we fit the data in
each interval with a cubic spline and subtract it from the
original time series to obtain the whitened residuals. 
The effect of whitening for different values of the filter width $W$
can be seen by calculating the autocorrelation function of the
whitened residuals.  The autocorrelation function measures the
similarity of a time series to itself upon translation in time by a
given lag $\tau$,
\begin{equation}
   ACF (\tau) = \sum_t \frac{(X_t - \mu)(X_{t+\tau} - \mu)}{\sigma^2}, 
\label{acfeq}
\end{equation}
where $X_t$ is the measured signal at time $t$, $\mu$ is the
mean of $X_t$, and $\sigma$ is the square root of the variance of
$X_t$.  In this example, for $W=300$ the polynomial has been removed
well, but significant periodicity remains in the time series (Fig.
\ref{synthetic}, 2nd panel).  For $W=150$, some of the periodicity has been
removed, but high frequency components are still apparent (Fig.
\ref{synthetic}, 3rd panel).  Using a filtering parameter $W=75$ removes
almost all of the wander in the time series, leaving a correlation
function consistent with Gaussian noise (Fig.
\ref{synthetic}, 4th panel). Recall that the highest-frequency sine wave in the
series has a period of 19.6, less than $W$, but the high-frequency 
components have been suppressed in proportion to the period.

\begin{figure}
{\includegraphics[scale=.80]{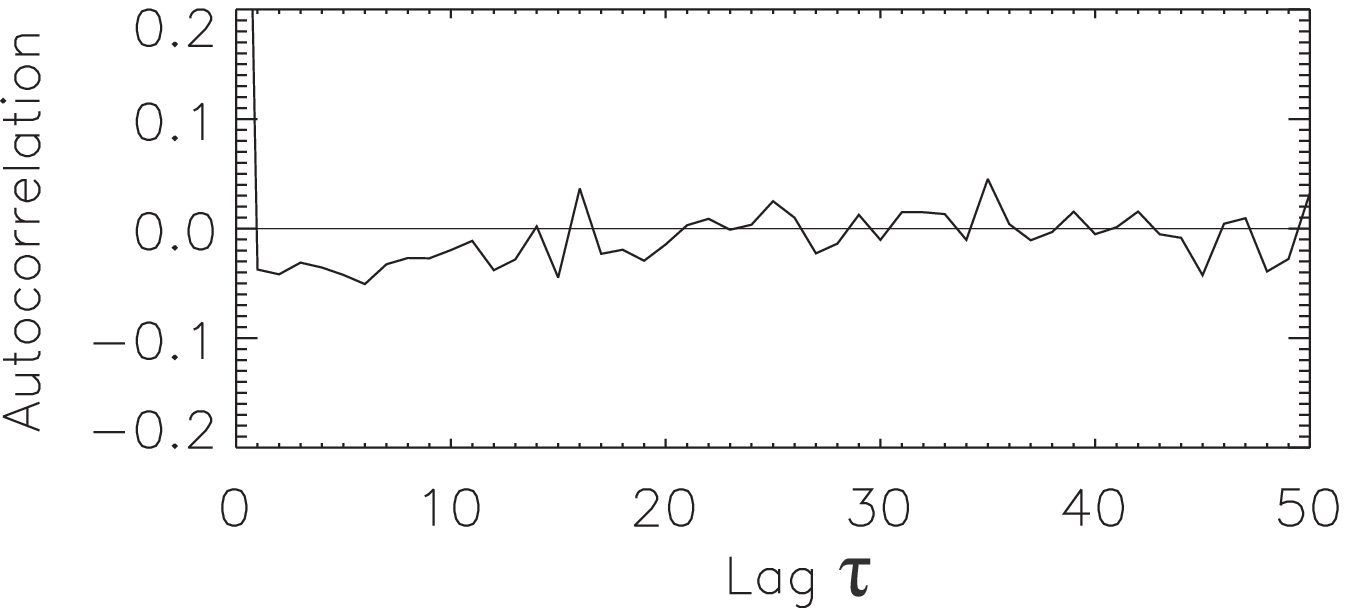}}
\caption{
Autocorrelation function of the simulated time series in Fig. \ref{synthetic}
(top panel), after whitening with $W=20$, showing
anti-correlation at low lag, the signature of overwhitening.  The
anti-correlation is strongest for lags $\sim W/2$.}
\label{overdt}
\end{figure}

Anti-correlations arise if $W$ is too small, comparable to the
sampling interval.  In this case, the cubic spline begins to fit the
noise, introducing spurious power at high frequencies. This results in
a whitened time series which is anti-correlated for lags $\tau < W$
(Fig. \ref{overdt}), as fitting the noise produces residuals which
are more likely to be of opposite sign.  To see this, consider the
limiting case in which each bin of width $W$ contains only two points.
The average value of each bin is subtracted from the time series,
resulting in data that are {\em anti}-correlated for a lag
$\tau=1$.  In general, the anti-correlation due to overwhitening is
strongest for a lag $\tau \simeq W/2$. This distinct signature allows
us to easily determine if the selected filter width $W$ is too small. As a
general rule, we need to keep $W$ at least several times larger than
the average sampling interval.

\subsection{Comparison of the Time-Domain Analysis with Fourier Methods}
 
\begin{figure}
{\includegraphics[scale=.80]{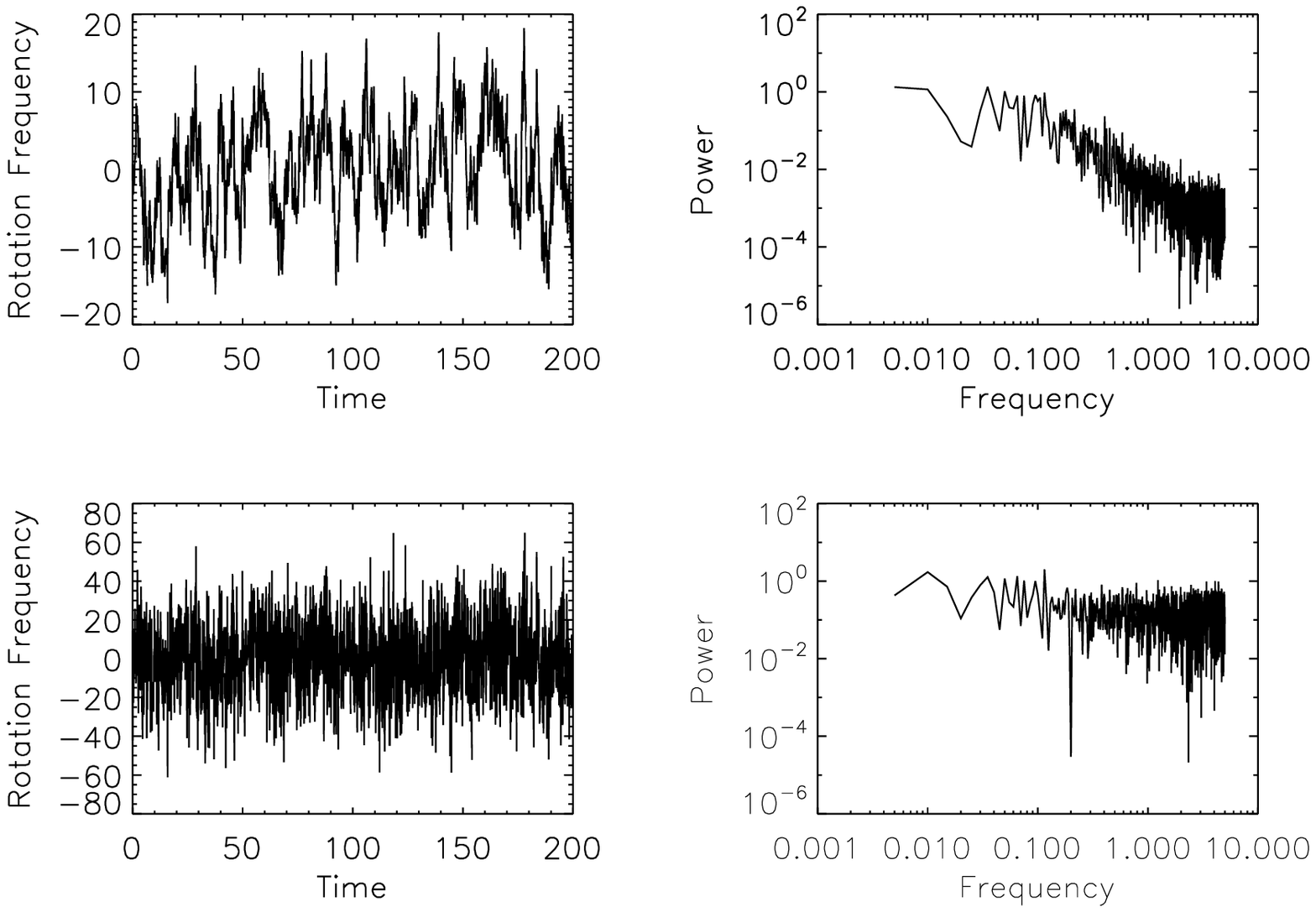}}
\caption{
{\sl Top left} - Simulated rotation frequency residuals with data
correlated over a timescale $\tau = 1$.  {\sl Top right} - The power
spectrum of the frequency residuals, indicating a ``knee" at $\omega
\tau_c \sim 1$. {\sl Bottom left} - The simulated residuals with Gaussian
noise. {\em Bottom right} - The power spectrum of simulated residuals.  After the
addition of noise to the time series, the coupling timescale is no
longer evident.}
\label{mysim1}
\end{figure}

To illustrate the advantages of time domain versus frequency domain
analysis for the type of time series that we analyze, we consider simulated
data produced by a simple physical model for a neutron star with an internal
degree of freedom.  The interaction between the fluid component and
crust component of the star is described by
\begin{equation}
   I_c \dot \Omega_c(t) = N(t) - \frac{I_c}{\tau_c} 
\left(\Omega_c(t) - \Omega_f(t)\right), 
\label{2comp3}
\end{equation}

\begin{equation}
   I_f \dot\Omega_f(t) = \frac{I_c}{\tau_c} \left(\Omega_c(t) - 
\Omega_f(t)\right), 
\label{2comp4}
\end{equation}
where $I_c$ and $I_f$ are the moments of inertia for the crust and
fluid, respectively, $\Omega_c(t)$ and $\Omega_f(t)$ are the rotation rates
of the crust and fluid, $\tau_c$ is the coupling timescale between the
crust and fluid, and $N(t)$ is the torque on the crust, from internal or
external sources.  To construct a simple model of neutron star spin
behavior, we consider crust and fluid components initially in
co-rotation, with a ratio of moments of inertial $I_c/I_f = 1$ for
illustration.  We perturb the crust with a series of
$\delta$-functions,
\begin{equation}
    N (t) = \sum_i A_i \delta (t-t_i),
\end{equation}
where $A_i$ are randomly generated amplitudes.  This model for the torque represents a series of instantaneous transfers of
angular momentum to the crust, regardless of the source or
sources.  In between impulses, eqs. (\ref{2comp3}) and (\ref{2comp4})
have the solution 
\begin{equation}
    \Omega_c (t) = \Omega_1 \frac{I_f}{I} e^{-t/\tau} + \Omega_2,
\end{equation}
where $\Omega_1$ and $\Omega_2$ are constants determined by $A_i$ of
the last $\delta$-function, $I\equiv I_f + I_c$ is the total moment of
inertia, and $\tau = (I_f/I_c) \tau_c$.  Using this model, we
construct a time series of frequency residuals shown in
Fig. \ref{mysim1}, consisting of 2000 points with an impulse applied
every unit time, and a sampling rate of 10 points per unit time.

\begin{figure}
\centering
{\includegraphics[scale=.70]{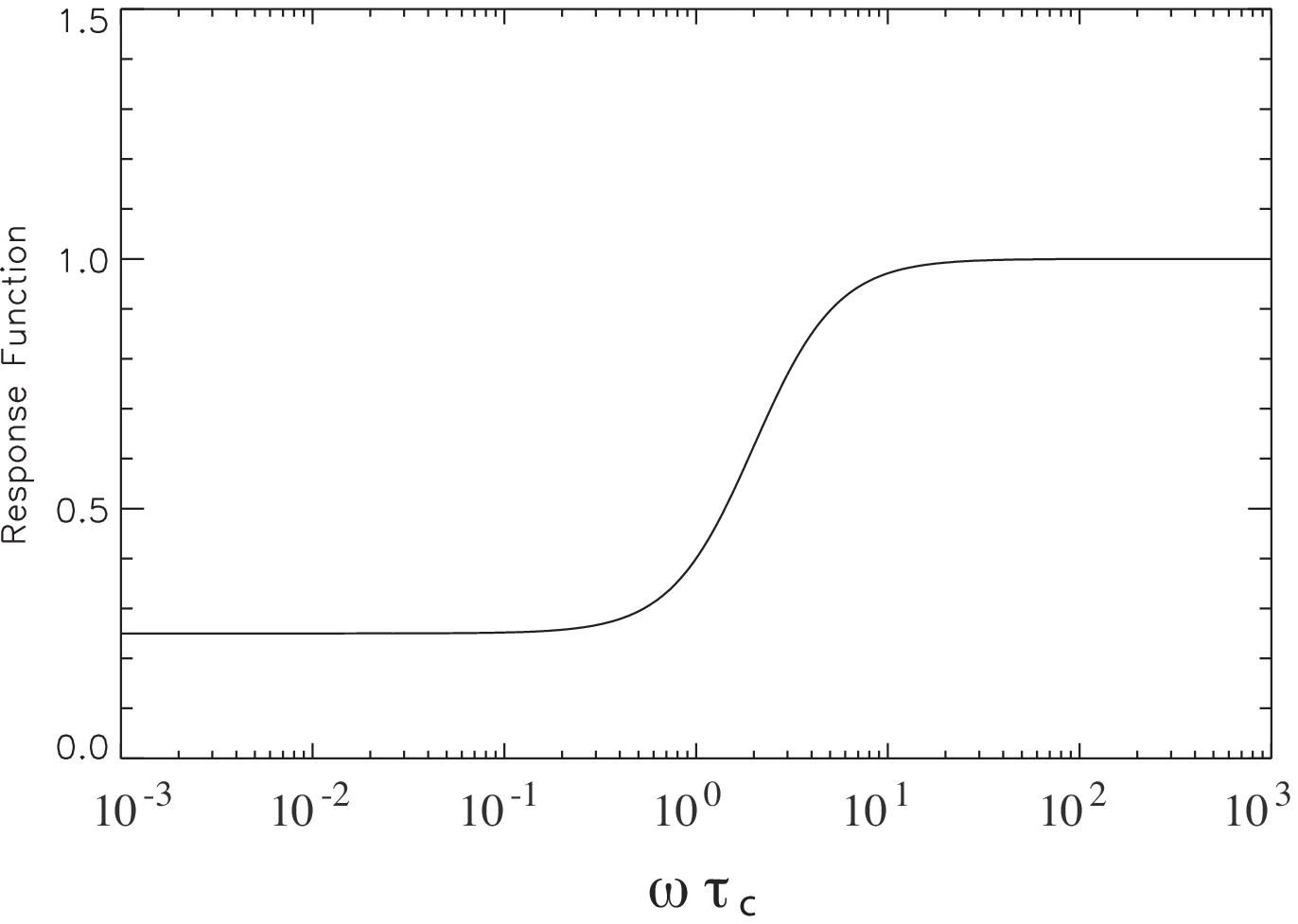}}
\caption{
Frequency dependent response function of the crust in the
two-component model, normalized to unity. 
The knee at $\omega\tau_c$ represents the transition from
coupled to decoupled response. 
}
\label{knee}
\end{figure}

The rapidly-perturbed system has relaxation response over a timescale
$\tau_c$, as can be seen from the response function in frequency.
 Fourier transforming eqs. (\ref{2comp3}) and (\ref{2comp4}), and
eliminating the Fourier transform of $\Omega_f(t)$, gives
\begin{equation}
    \left | \tilde{\Omega}_c (\omega) \right | ^2 = 
\frac{1}{\omega^2 I_c^2} 
\left [ \frac{(\omega \tau_c)^2 + (I_c/I_f)^2}{(\omega \tau_c)^2 +
(I/I_f)^2} \right ] \left | \tilde{N}(\omega) \right |^2;
 \label{response}
\end{equation}
We define the response function as the
frequency dependent term in brackets, shown in Fig. \ref{knee}. 
At frequencies higher than $\omega \tau_c = 1$, only the crust
responds to the torque.  At lower frequencies, the entire
moment of inertia of the star $I$ is perturbed by the torque,
resulting in a lower response. There is 
a ``knee" in the spectrum at $\omega \tau_c \simeq 1$.
Such a knee is evident in the power spectrum of the simulated phase
residuals (Fig. \ref{mysim1} - top right panel).

\begin{figure}
{\includegraphics[scale=.60]{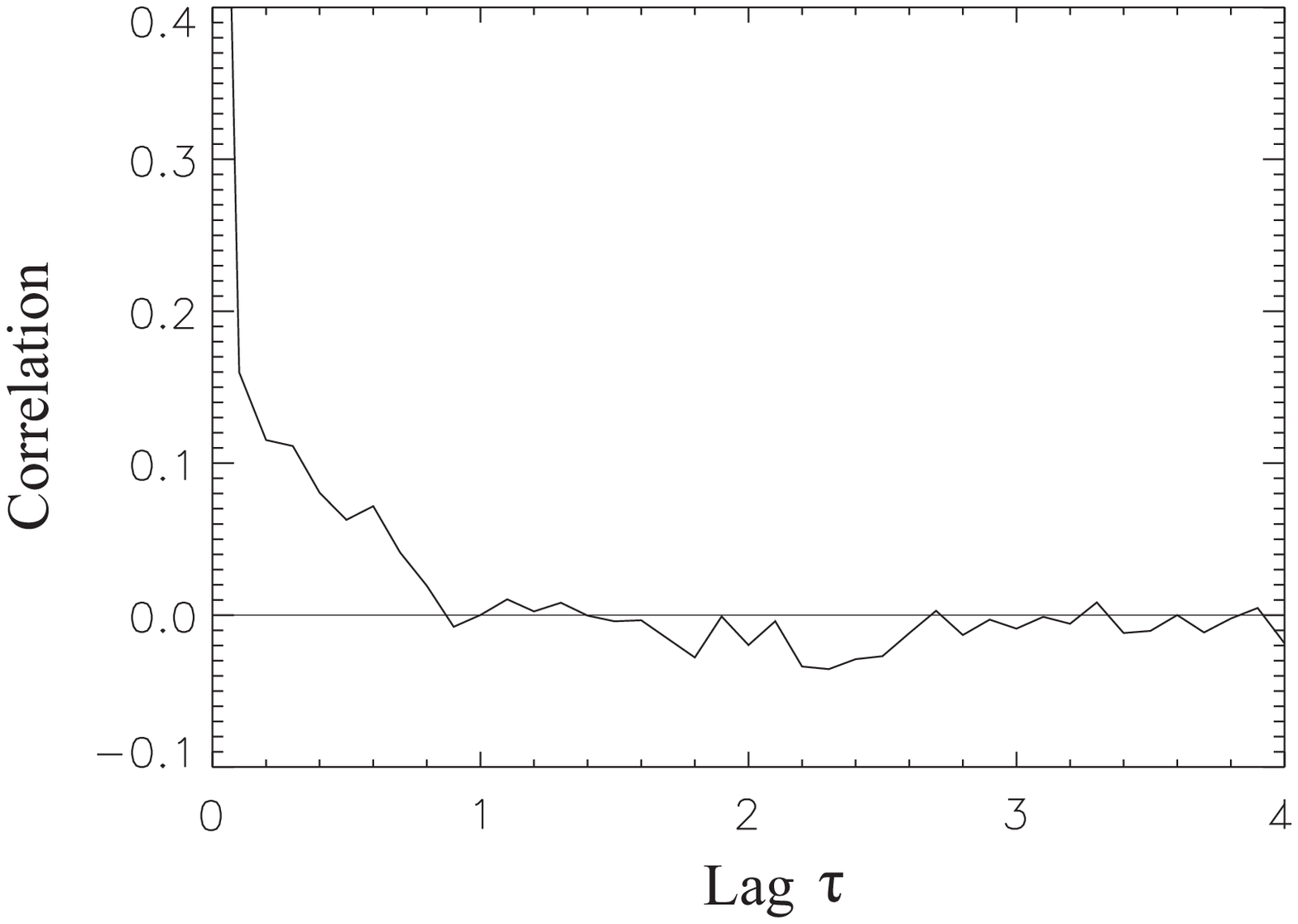}}
\caption{The autocorrelation function for the simulated time series shown in Fig. \ref{mysim1} (bottom left panel), showing the intrinsic relaxation time of the system that is not visible in the power spectrum.} 
\label{mysim3}
\end{figure}

In Fig. \ref{mysim1} (bottom panels), we have added gaussian noise to the time
series to simulate noisy data.  In this case, the knee in the spectrum
corresponding to the coupling timescale $\tau_c$ becomes buried beneath the
noise.  Time domain analysis is better suited for this noise-dominated
time series.  We calculate the autocorrelation function of the time
series shown in Fig. \ref{mysim1} (bottom left).  The coupling time is
readily seen in the autocorrelation function shown in
Fig. \ref{mysim3}.

\subsection{DCF Tests}

\begin{figure}
\centering

{\includegraphics[scale=.6]{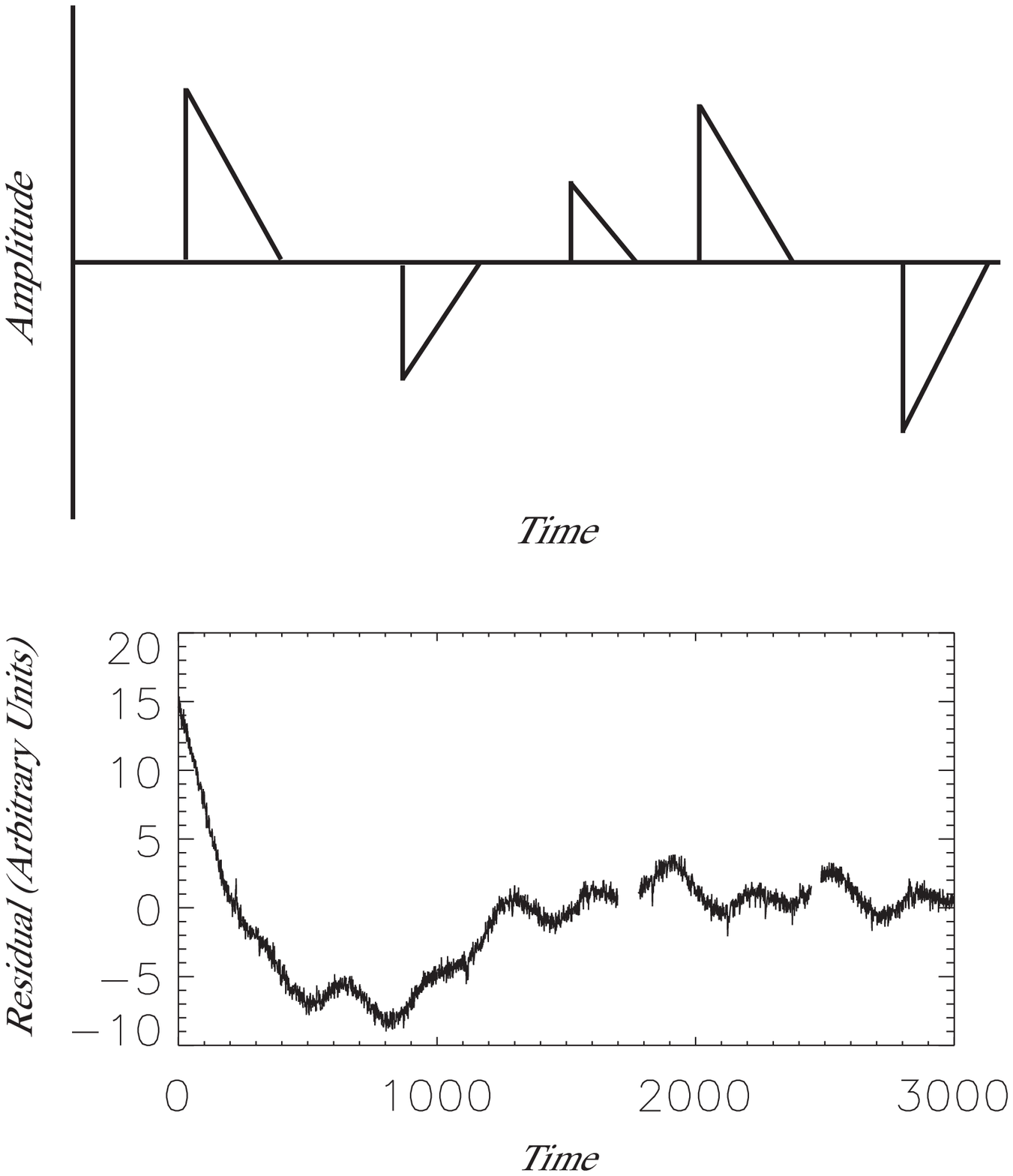}}
\caption{
{\sl Top} - Schematic of the sawtooth function.  Each impulse decays
linearly over a timescale of 10 units.  Impulses are randomly spaced
in time, with random amplitudes of either sign drawn from a Gaussian
distribution with a standard deviation of unity. {\sl Bottom} -
Simulated residuals consisting of periodic functions, a polynomial,
Gaussian noise, and a sawtooth function.  Several gaps are added to
simulate the sampling we have for the real data.  The effects of the sawtooth function are too
small to be seen against the wander and added noise.}
\label{sawpic}
\end{figure}

To illustrate that the DCF can distinguish between correlations due to
wander and those due to a relaxation process, we add a series of 10
``sawtooth" functions to the synthetic data of Fig.
\ref{mysim1} (lower panel). 
Each sawtooth consists of a discontinuous jump of random
magnitude, followed by a linear decay over 10 time units.  The jumps
are randomly spaced over the time series, with magnitudes drawn from a
Gaussian with a standard deviation of unity about zero; see 
Fig. \ref{sawpic} (top panel). The resulting data are shown in Fig.
\ref{sawpic} (bottom panel); the effects of the sawtooth are too small
to be seen in these simulated data. We have also added gaps to simulate real data.  We show data with even sampling (apart from the gaps); we have confirmed that uneven sampling has a negligible effect on the results.

\begin{figure}
{\includegraphics[scale=1.0]{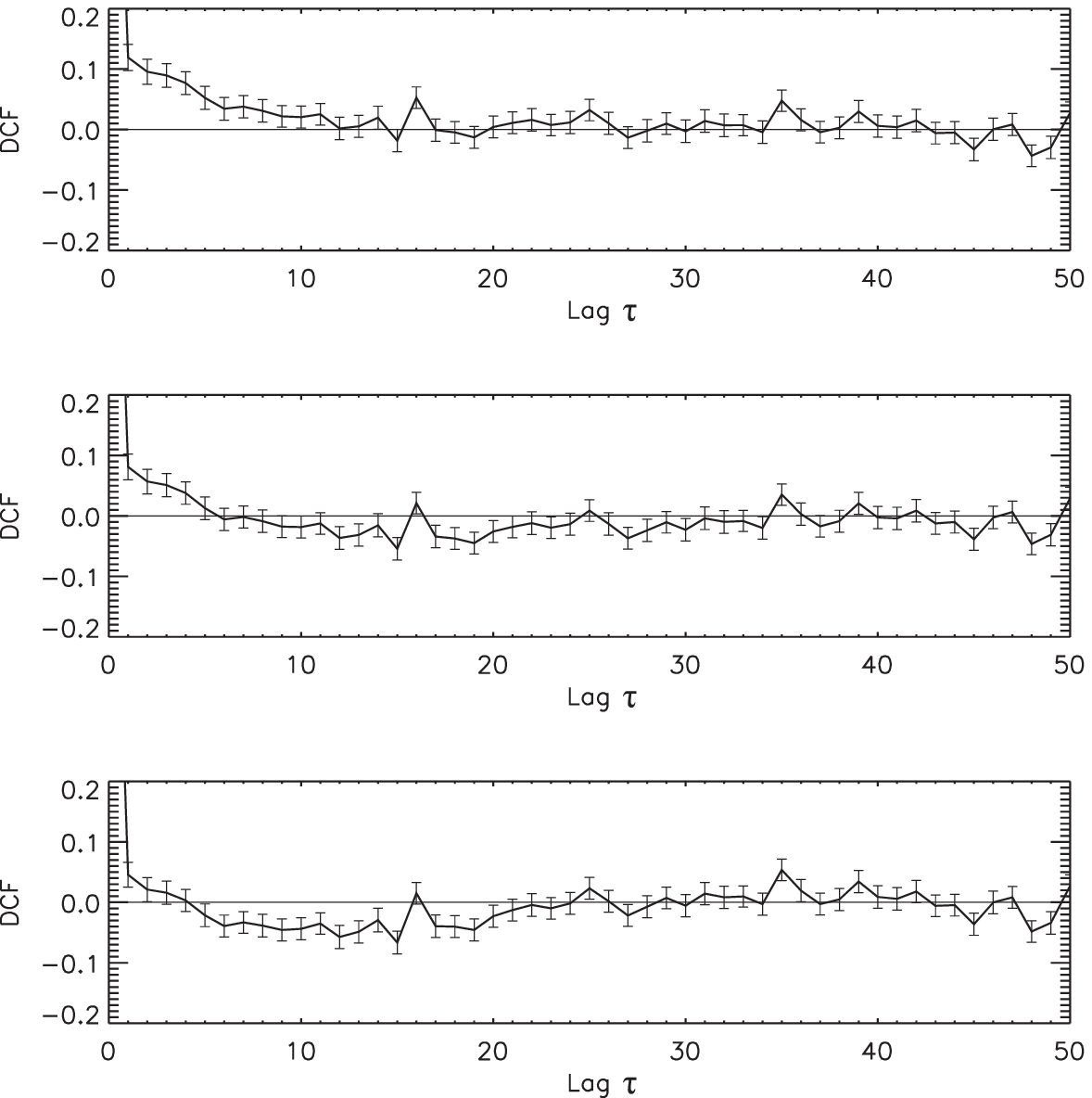}}
\caption{
DCFs for whitened simulated residuals using $W=75,50,25$, after adding
a ``sawtooth" function to the simulated residuals shown in Fig.
\ref{sawpic}.  The relaxation timescale of 10 units is easily
identified for $W=75$, with low-frequency wander nearly completely
subtracted by the high-pass filter.  For $W=50$, the fitting function
begins to subtract the correlations.  At $W=25$, the time series is
overwhitened, resulting in reduction of the relaxation signature and
anticorrelations from lags in the range $5 \lesssim \tau \lesssim 25$.}
\label{sawdcf}
\end{figure}

DCFs for this time series upon whitening are shown in
Fig. \ref{sawdcf} for three filter widths $W$.  For $W=75$, only
positive correlations over $\sim 10$ time units are evident (top
panel), as the low-frequency wander has been almost completely
subtracted.  This DCF signature is easily differentiated from the
variability arising from the impulsive noise, which produces both
correlations and anti-correlations.  At lower values of $W$, the
fitting function begins to remove correlations in the data, and
anti-correlations begin to appear (middle and bottom panels); these
data have been overwhitened.  For a broad range in $W$, correlations
remain evident, even if the data have been overwhitened.

\begin{figure}
\includegraphics[scale=1.0]{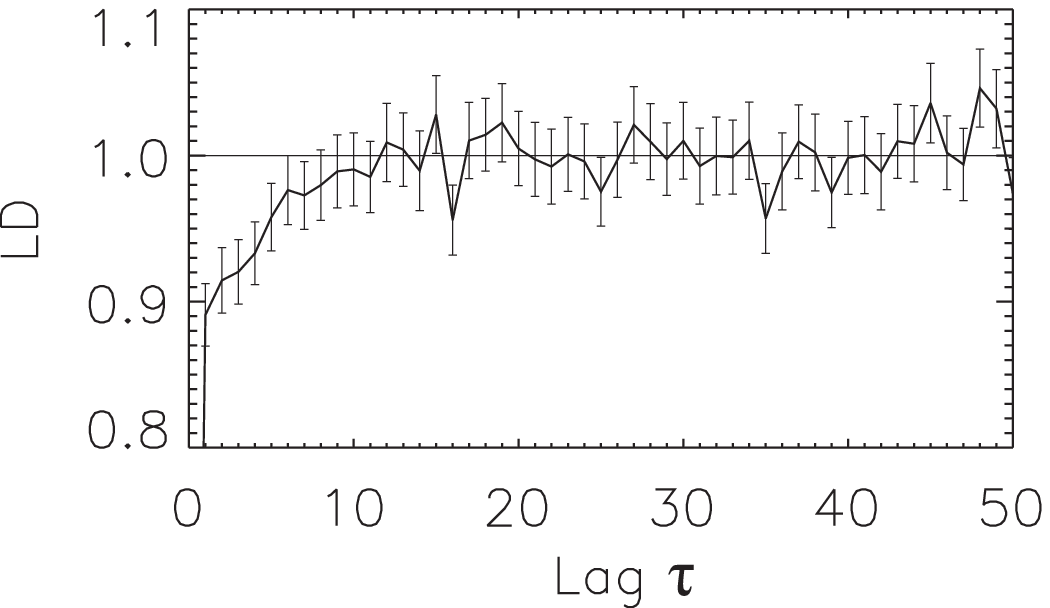}
\caption{LD of the simulated time series shown in Fig. \ref{sawpic}, after whitening with $W=75$.  The relaxation process occurring over 10 time units is evident, confirming the results of the DCF shown in Fig. \ref{sawdcf}.}
\label{sawld}
\end{figure}

The LD for the synthetic data of Fig. \ref{sawpic}
(bottom panel) is shown in Fig. \ref{sawld}.  We decorrelated the data
through random shuffling of the points. For 100 such shufflings, we
calculated the average ${\rm LD}(\tau)$ to establish the base ${\rm LD}(\tau)$ one
would expect under the null hypothesis that the data are completely
uncorrelated. These values are shown as the horizontal lines in each
figure; ${\rm LD}(\tau)$ is normalized in terms of this average value. This
method has the advantage that nothing is assumed about the underlying
statistics of the data, rather, the data themselves are used to
evaluate ${\rm LD}(\tau)$ for the null hypothesis. Uncertainties are
calculated using eq. \ref{error}.  The LD shows that the data are
correlated over a timescale of 10 time units, consistent with results
of the DCF for this simulated time series (Fig.
\ref{sawdcf}).

We conclude that high-pass filtering of the time series followed by
calculation of the DCF is a robust method for identifying an intrinsic
relaxation timescale $\tau_c$, provided the following conditions are
met:
\begin{equation}
   \Delta t_{\rm samp} < \tau_c < \tau_{\rm wander},
\end{equation}
where $\Delta t_{\rm samp}$ is the mean sampling interval and
$\tau_{\rm wander}$ is the shortest timescale of the wander.  If the first
inequality is not satisfied, then the time resolution of the data is
not sufficient to resolve the correlation timescale.  If the second
condition is not met, then the correlation cannot be disentangled from
the wander.  In practice, we must also require
\begin{equation}
\tau_{\rm corr}<W<\tau_{\rm wander}, 
\end{equation}
to ensure that the filter removes the wander but not the
correlation. In the extreme case of $W\sim \tau_{\rm samp}$, the
filtering method introduces spurious {\rm anti}-correlations,
indicating that $W$ is too small. 

\subsection{Robustness Tests}

\begin{figure}
\centering
\includegraphics[scale=.8]{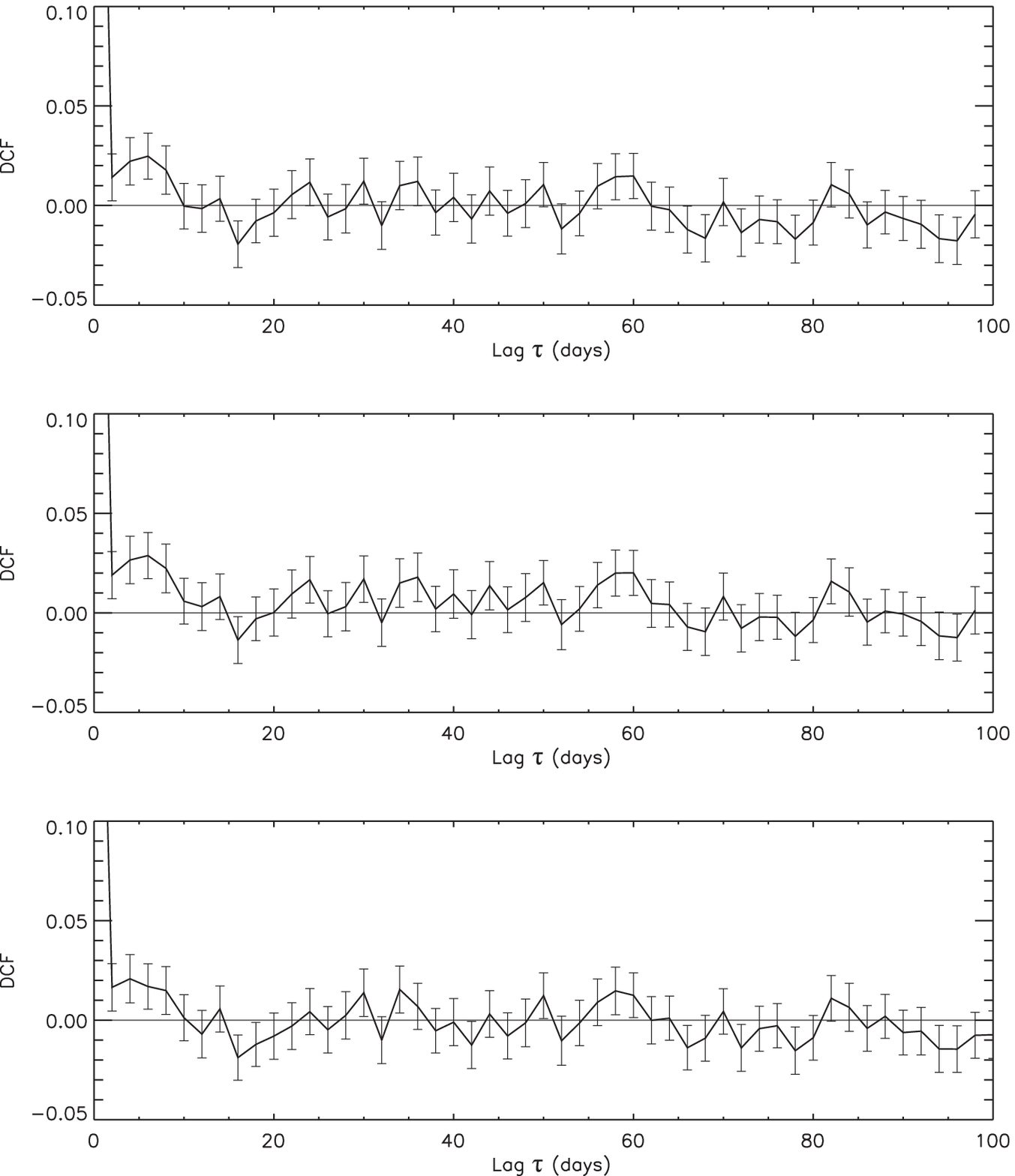}
\caption{
DCF for PSR B1133+16, using three different whitening methods.  In the
top panel, the whitening method described in Section 2.1 is used with
$W=400$ d. (For other values of $W$, see Fig. \ref{1133w}).  For the
middle panel, timing residuals were whitened using a fifth-order
polynomial.  For the bottom panel, the DCF function itself was
whitened using $W=400$ d (see text).}
\label{dcf3}
\end{figure}

To verify that the correlations found using the DCF are not introduced
by the whitening method, we compare our results to those found using
other whitening methods, using data from PSR B1133+16 for
illustration.  The wander in this pulsar can be adequately removed
with a fifth-order polynomial.  Applying the DCF analysis to the data
whitened using this technique, the results are very similar to those
we find using the technique described in Section 2.1 (Fig. \ref{dcf3}
middle panel).  We also apply the DCF to unwhitened data for PSR
B1133+16, and then whiten the correlation function, rather than the
time series, to remove the signature of low-frequency wander from the
DCF (Fig. \ref{dcf3}, bottom panel).  These three whitening techniques
give very similar results.

\begin{figure}
{\includegraphics[scale=.9]{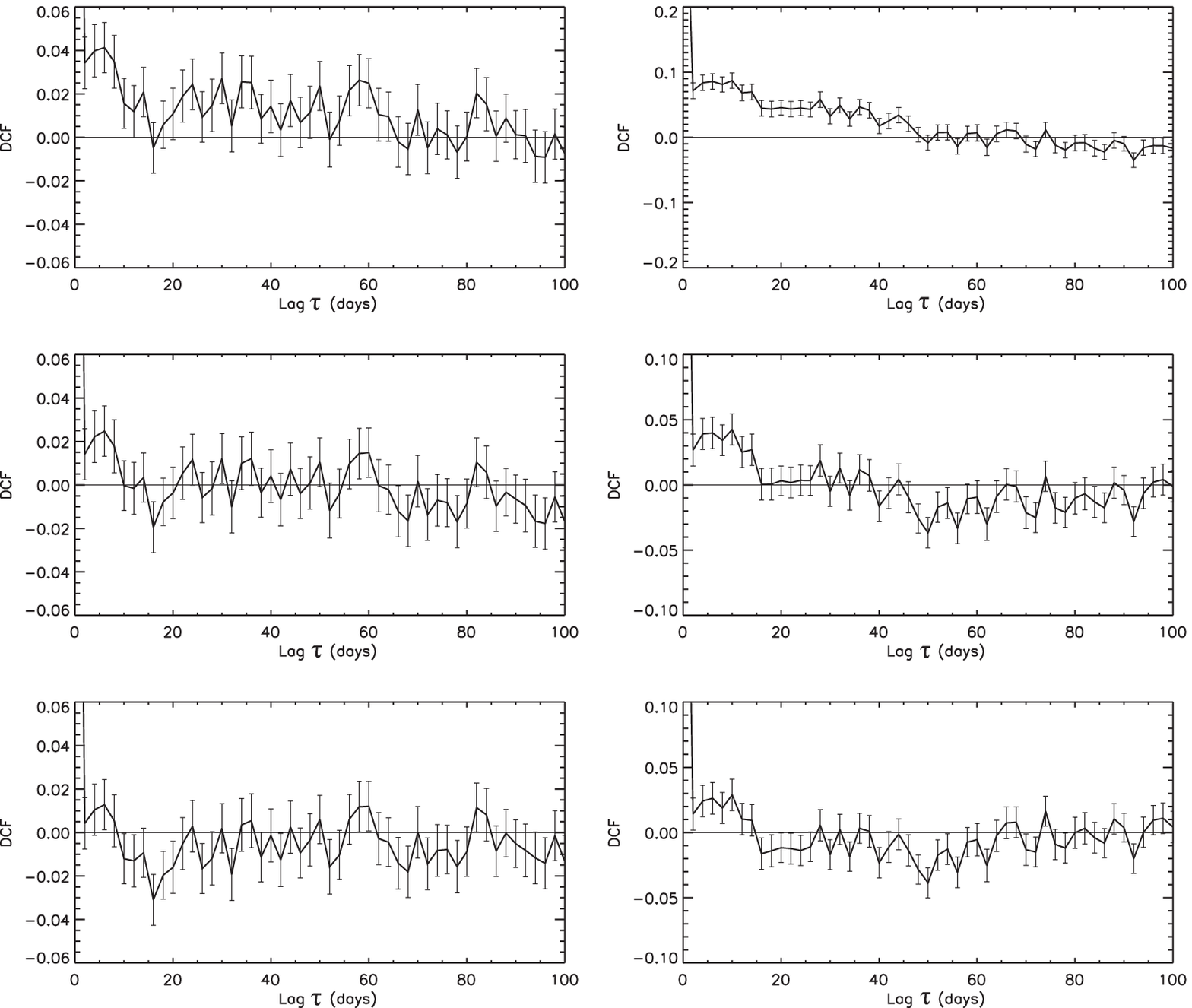}}
\caption{
DCFs for PSR B1133+16 (left column) and PSR B1933+16 (right column)
for several values of $W$.  For B1133+16, the top panel uses $W=600$
days, the middle panel uses $W=200$ days, and the bottom panel uses
$W=80$ days.  In the range $W \sim 100-500$ days, the DCFs are nearly
indistinguishable.  For B1933+16, the top panel uses $W=140$ days, the
middle panel uses $W=100$ days, and the bottom panel uses $W=60$ days.
For $W \sim 80-120$ days, the DCFs are similar and show correlations
for $\tau\lesssim 30$ d.  }
\label{1133w}
\end{figure}

For B1133+16, the large difference between the relaxation timescale
($\sim 10$ d) and the wander timescale ($\sim 1000$ d) allows a wide
range of choices for the high-pass filter width $W$ that give similar
results.  The wander is not well removed for $W \gtrsim 600$ days, and
the fitting function begins to subtract the relaxation correlations
for $W\lesssim 100$ days.  We show DCFs for PSR B1133+16 for several
values of $W$ in Fig.
\ref{1133w} (left column).

The characteristic timescale of the wander is much shorter for
B1933+16, about 300 d.  We plot several DCFS for B1933+16 for
different values of $W$ (Fig. \ref{1133w}, right column).  To
successfully remove the wander, $W$ must be a factor of several
smaller than the shortest characteristic timescales of the wander.  We
also ensure that $W$ is a factor of 2-3 greater than the relaxation
timescale that appears in the DCF as $W$ is reduced. In the case of
B1933+16, these constraints leave little room to vary $W$.  For $W \sim 80-120$
days, the DCFs are nearly identical.  For Figs. \ref{resids1} and
\ref{resids2}, we use a cutoff frequency of $f\sim 400$ d$^{-1}$ for
B1133+16, and $f\sim 120$ d$^{-1}$ for B1933+16.

\end{document}